\documentclass[manuscript]{acmart}
\AtBeginDocument{%
  }

\setcopyright{acmlicensed}
\copyrightyear{2026}
\acmYear{2026}
\acmDOI{XXXXXXX.XXXXXXX}
\acmISBN{978-1-4503-XXXX-X/2018/06}

\usepackage{xspace}
\usepackage{enumitem}
\usepackage{tabularx}
\usepackage{multicol}
\usepackage{multirow}
\usepackage{amsmath}
\usepackage{graphicx}
\usepackage{algorithmic}
\usepackage{algorithm}
\usepackage{booktabs}
\usepackage{makecell}
\usepackage{caption}
\usepackage{colortbl}
\usepackage{tikz}
\definecolor{mytitle}{rgb}{0.9, 0.9, 0.9}
\definecolor{mygray}{rgb}{0.95, 0.95, 0.95}
\usepackage{subcaption}
\usepackage{booktabs}
\usepackage{multirow}
\usepackage{graphicx}
\usepackage{pifont}

\usepackage{booktabs}
\usepackage{multirow}
\usepackage{caption}

\usepackage{booktabs}
\usepackage{xcolor}
\usepackage{tikz}
\usepackage{graphicx}

\newcommand{\rcbar}[2]{%
  \pgfmathsetmacro{\pct}{100*(#1)/(#2)}%
  \pgfmathsetmacro{\barw}{0.85*(#1)/(#2)}%
  \begin{tabular}{@{}c@{}}
    \begin{tikzpicture}
      \fill[gray!18] (0,0) rectangle (0.85,0.13);
      \fill[blue!55] (0,0) rectangle (\barw,0.13);
    \end{tikzpicture}\\[-0.2em]
{\scriptsize #1/#2 (\pgfmathprintnumber[fixed,precision=1]{\pct}\%)}
  \end{tabular}%
}

\usetikzlibrary{arrows.meta,positioning,fit,calc,backgrounds,shapes.geometric}
\usepackage{xcolor}
\usepackage{listings}
\usepackage{listings}
\usepackage{xcolor}
\usepackage{tcolorbox}
\tcbuselibrary{breakable}
\newtcolorbox{resultbox}{
    colback=gray!15,
    colframe=black!70,
    boxrule=0.6pt,
    arc=1mm,
    left=1mm,
    right=1mm,
    top=0.5mm,
    bottom=0.5mm,
    breakable
}

\lstdefinelanguage{Solidity}{
  sensitive=true,
  morekeywords={
    pragma,solidity,contract,interface,library,abstract,
    function,modifier,event,constructor,receive,fallback,
    public,private,internal,external,payable,view,pure,
    memory,storage,calldata,returns,return,
    mapping,address,bool,string,bytes,uint,uint256,int,int256,
    if,else,for,while,do,break,continue,require,revert,
    emit,assembly,new,is,import,from
  },
  morecomment=[l]{//},
  morecomment=[s]{/*}{*/},
  morestring=[b]"
}

\lstdefinestyle{solidity}{
  language=Solidity,
  basicstyle=\ttfamily\footnotesize,
  keywordstyle=\color{blue!70!black}\bfseries,
  commentstyle=\color{gray}\itshape,
  stringstyle=\color{green!40!black},
  frame=single,
  rulecolor=\color{black!35},
  numbers=left,
  numberstyle=\tiny\color{gray},
  numbersep=6pt,
  breaklines=true,
  columns=fullflexible,
  keepspaces=true,
  showstringspaces=false,
  tabsize=2,
  captionpos=b
}

\begin{document}

\title{Price is Only the Symptom: Understanding and Analyzing Economic-State Manipulation Vulnerabilities in DeFi}


\author{Hao Wu}
\affiliation{%
  \institution{Xi'an Jiaotong University}
  \city{Xi'an}
  \country{China}}
  \affiliation{%
  \institution{The Hong Kong Polytechnic University}
  \city{Hong Kong}
  \country{China}
}
  
\email{emmanuel\_wh@stu.xjtu.edu.cn}

\author{Haijun Wang}
\authornote{Corresponding author.}
\affiliation{%
  \institution{Xi'an Jiaotong University}
  \city{Xi'an}
  \country{China}
}
\email{haijunwang@xjtu.edu.cn}

\author{Xiapu Luo}
\authornotemark[1]
\affiliation{%
 \institution{The Hong Kong Polytechnic University}
 \city{Hong Kong}
 \country{China}}
 \email{csxluo@comp.polyu.edu.hk}

\author{Ming Fan}
\affiliation{%
  \institution{Xi'an Jiaotong University}
  \city{Xi'an}
  \country{China}}
\email{mingfan@mail.xjtu.edu.cn}

\author{Ting Liu}
\affiliation{%
  \institution{Xi'an Jiaotong University}
  \city{Xi'an}
  \country{China}}
\email{tingliu@mail.xjtu.edu.cn}

\renewcommand{\shortauthors}{Hao Wu, Haijun Wang, Xiapu Luo, Ming Fan, and Ting Liu}

\begin{abstract}
Economic-state manipulation has emerged as a significant security threat to decentralized finance (DeFi), causing substantial financial losses. Existing studies predominantly characterize these attacks as price manipulation, focusing on observable price distortions while overlooking underlying violations of economic-state integrity. We conduct a comprehensive empirical study of Economic-State Manipulation (ESM) vulnerabilities in DeFi. We first curate a diverse dataset comprising 276 real-world attacks and 91 Code4rena audit findings. Based on these cases, we develop a cause-oriented taxonomy comprising four primary categories and eleven subcategories. Our analysis reveals that ESM vulnerabilities extend beyond conventional price manipulation, frequently involving violations of protocol-internal accounting, AMM state consistency, and execution-time constraints. We further investigate the evaluation datasets of six existing studies and find that, although they cover 6–10 vulnerability subcategories, their distributions remain dominated by limited manipulation mechanisms. Furthermore, we systematically evaluate seven representative detection approaches on our dataset. Our results reveal limited coverage, with the best-performing approach identifying only 50.4\% of real-world attacks. Through an analysis of false negatives and false positives, we identify fundamental limitations arising from reliance on predefined attack patterns, insufficient reasoning about economic-state semantics, and incomplete exploitability validation. These findings provide empirical insights for developing more generalizable ESM vulnerability detection techniques.

\end{abstract}

\maketitle

\section{Introduction}
Decentralized Finance (DeFi) has emerged as one of the most prominent applications of blockchain ecosystem, enabling financial services such as token exchange, lending, and flash loans. 
DeFi protocols are typically composed of a series of smart contracts, which are self-executing programs deployed on the blockchain, collectively managing substantial amounts of digital assets.
However, since smart contracts are publicly accessible on the blockchain, potential vulnerabilities within DeFi protocols can be discovered and exploited by hackers. In 2025 alone, security incidents related to smart contracts led to losses of about \$611.3 million~\cite{web3security}.

DeFi protocols rely heavily on economically meaningful states to perform critical financial operations, including asset valuation, exchange execution, and reward distribution. These economic states, such as asset prices, exchange rates, and share values, serve as the foundation for protocol decision-making. However, unlike traditional financial systems, DeFi protocols expose these states through programmable and composable mechanisms, making them vulnerable to adversarial manipulation. By intentionally influencing economic states consumed by victim protocols, attackers can force protocols to operate under adversarially constructed conditions and obtain significant illicit profits. Such attacks have become one of the most severe security threats in the DeFi ecosystem. A statistical analysis of 276 real-world attacks collected from DeFiHackLabs~\cite{DeFiHackLabs} and exsiting datasets~\cite{wu2023defiranger,kong2023defitainter,zhong2025detecting,xie2024defort,zhang2025following,liu2025llm}, spanning from October 2020 to March 2026, reveals cumulative losses of \$524 million USD. These incidents demonstrate that economic-state manipulation represents a persistent and financially devastating threat to DeFi security. 

Despite their significant impact, existing discussions and analyses of these attacks have largely focused on the manipulated price as the primary indicator, leading to the widespread characterization of them as Price Manipulation (PM) attacks. However, price manipulation captures only a subset of a broader phenomenon. In practice, attackers may manipulate various protocol-relevant economic states beyond explicit price values, such as liquidity reserves in automated market makers or exchange rates in lending protocols. These manipulated states are subsequently consumed by security-critical operations, causing protocols to make incorrect financial decisions.
Therefore, we argue that price manipulation is only a symptom of a broader class of vulnerabilities, which we refer to as \textbf{Economic-State Manipulation (ESM) attacks}. ESM attacks describe scenarios where adversaries intentionally manipulate economically meaningful states or intermediate values before they are consumed by security-critical operations, causing protocols to execute under adversarially constructed economic conditions. This perspective shifts the focus from the observable consequence (e.g., abnormal price changes) to the underlying security failure: the protocol's incorrect trust in manipulable economic states.

\begin{table*}[t]
    \centering
    \caption{Top 10 DeFi ESM Attack Incidents by Loss (2025-2026)}
    \vspace{-0.2cm}
    \label{tab:recent-severe-incidents}
    \footnotesize
    \setlength{\tabcolsep}{2.5pt}       
    \renewcommand{\arraystretch}{0.88} 
    \begin{tabular*}{0.94\textwidth}{
        @{\extracolsep{\fill}}
        c r l c
        c r l c
        @{}
    }
        \hline
        Incident & Loss & Chain & Time &
        Incident & Loss & Chain & Time \\
        \hline

        DCFToken
            & 442k
            & BSC
            & 2025-03-18
            & NGP
            & 2M
            & BSC
            & 2025-09-18 \\

        \shortstack[l]{Nalakuvara}
            & 105.5K
            & BASE
            & 2025-05-09
            & Moonwell
            & 1M
            & BASE
            & 2025-11-04 \\

        ResupplyFi
            & 9.6M
            & ETH
            & 2025-06-26
            & Makina
            & 5.1M
            & ETH
            & 2026-01-20 \\

        GMX
            & 42M
            & Arbitrum
            & 2025-07-09
            & Moonwell-2
            & 1.8M
            & BASE
            & 2026-02-15 \\

        d3xai
            & 159.5k
            & BSC
            & 2025-08-16
            & Curve LlamaLend
            & 240k
            & ETH
            & 2026-03-02 \\

        \hline
    \end{tabular*}
    \vspace{-0.2cm}
\end{table*}

As shown in Table~\ref{tab:recent-severe-incidents}, ESM attacks have remained prevalent in 2025-2026, causing substantial financial losses. The persistence and diversity of these incidents highlight the need to understand not only observable price distortions, but also the underlying economic states and assumptions that enable such attacks.
Early efforts, such as DeFiRanger~\cite{wu2023defiranger} and DeFort~\cite{xie2024defort}, focus on identifying price manipulation patterns from transaction-level behaviors. However, these approaches are limited in their ability to proactively prevent PM attacks. To address this limitation, tools such as DeFiTainter~\cite{kong2023defitainter} and SmartCat~\cite{zhang2025following} leverage predefined rules and static analysis to detect PM vulnerabilities prior to exploitation. 
More recently, advances in large language models (LLMs) have inspired hybrid approaches, where methods like GPTScan~\cite{sun2024gptscan} and PMDetector~\cite{liu2025llm} integrate LLMs with static analysis to enhance semantic understanding and improve detection capabilities. 
However, these approaches inherit a fundamental limitation from the existing PM-centric understanding: they primarily identify specific manipulation patterns rather than systematically analyzing the economic states being manipulated and the assumptions violated by victim protocols. As a result, existing techniques may fail to capture emerging attack variants where attackers manipulate non-price economic states or exploit complex dependencies between economic states and protocol logic.

The continuous emergence of ESM attacks raises fundamental questions regrading our current understanding and defense capabilities of ESM attacks. However, prior empirical studies have largely focused on general smart contract vulnerabilities or specific bug classes (e.g., reentrancy and access control), with limited attention to price-related manipulation attacks. Specifically, three fundamental gaps remain to be addressed.
(1) \textbf{Lack of a comprehensive taxonomy of ESM attacks}. Existing studies mainly focus on specific attack mechanisms or detection scenarios, without characterizing the underlying root causes of these attacks. (2) \textbf{Limited understanding of dataset representativeness}. 
Existing studies are often evaluated on narrow datasets, which may fail to capture the diversity of real-world ESM attacks. The distribution of ESM root causes in existing evaluation datasets remains insufficiently examined, potentially biasing evaluations toward prevalent manipulation patterns.
(3) \textbf{Insufficient evaluation of existing detection approaches}. Existing evaluations are often conducted on limited benchmarks or narrowly defined attack scenarios. Moreover, the lack of comprehensive and diverse evaluation datasets may lead to biased conclusions regarding their detection capability.

To fill this gap, we conduct the first comprehensive empirical study of ESM attacks in DeFi. Specifically, we curate a dataset of 177 real-world ESM attacks from DeFiHackLabs~\cite{DeFiHackLabs}, spanning from Oct 2020 to Mar 2026. We further incorporate all publicly available attack incidents from existing ESM/PM datasets~\cite{wu2023defiranger,kong2023defitainter,zhong2025detecting,xie2024defort,zhang2025following,liu2025llm}, adding 99 events after deduplication and manual validation.
In addition, we analyze 1283 peer-reviewed audit report findings from Code4rena~\cite{Code4rena} published between Jan 2021 and Mar 2026, identifying 91 audit findings related to ESM vulnerabilities. Our dataset substantially exceeds those used in prior detection studies (e.g., 95 PM attacks in DeFiScope~\cite{zhong2025detecting}), enabling a more thorough analysis. Based on this dataset, we aim to answer the following three research questions.

\textbf{RQ1: Taxonomy and Landscape of ESM Vulnerabilities.} What are the major root causes of ESM vulnerabilities observed in practice, and how are they distributed across real-world exploits and proactively discovered vulnerabilities? To answer RQ1, we construct a cause-oriented taxonomy of ESM vulnerabilities based on the violated economic integrity assumptions during potential exploitation. Our taxonomy identifies four primary categories and eleven subcategories, covering failures in \textit{Price-Source Trust}, \textit{Protocol-Internal Accounting}, \textit{AMM and Liquidity State Integrity}, and \textit{Execution-Time Protection}. There results demonstrate that ESM vulnerabilities extend beyond conventional price manipulation, with price distortion often emerging only as a downstream symptom of deeper economic-state manipulation.

\textbf{RQ2: Representativeness of Existing Evaluation Datasets.} To what extent do existing PM/ESM evaluation datasets cover the diversity of root causes observed in real-world ESM vulnerabilities? To answer RQ2, we map the evaluation datasets of six existing PM/ESM studies to our taxonomy and examine their coverage of different ESM root causes. We find that these datasets cover 6-10 of the 11 identified subcategories, indicating that recent benchmarks already include a relatively diverse set of ESM vulnerability mechanisms. However, it exhibits uneven distributions with many cases concentrated in a few manipulation mechanisms. More importantly, the presence of diverse vulnerability instances in a benchmark does not necessarily imply that the corresponding detection approach can recognize their underlying root causes. Existing techniques are largely designed around observable symptoms or predefined attack patterns rather than the underlying economic-state violations. This findings motivates a root-cause level evaluation of existing detection techniques in RQ3.

\textbf{RQ3: Effectiveness and Limitations of Existing Detection Approaches.} How effectively do state-of-the-art approaches detect diverse ESM vulnerabilities, and what fundamental limitations lead to missed detections and false positives? To answer RQ3, we evaluate seven representative approaches spanning transaction-dependent analysis, bytecode program analysis, and LLM-assisted sourcecode analysis. We find that existing methods provide only partial coverage of real-world ESM vulnerabilities, with the best-performing approach detecting 50.4\% of the evaluated attacks. Our analysis further reveals a fundamental trade-off across detection paradigms. Transaction-dependent methods rely on observable execution evidence and known attack behaviors, whereas proactive program analysis approaches remain constrained by predefined patterns, taint abstraction, and candidate-generation mechanisms. Moreover, false positives in proactive approaches mainly stem from incomplete exploitability validation, such as failing to determine whether a suspicious state is attacker-controllable. Overall, existing techniques lack systematic reasoning about the underlying mechanisms that make ESM vulnerabilities exploitable, including which economic states are manipulated, which integrity assumptions are violated, and how such violations propagate to security-critical operations. 

In summary, the contributions of this paper are as follows. All datasets and evaluation code are released in~\cite{Evaluation}.

\begin{itemize}[leftmargin=*]
    \item We curate a comprehensive ESM dataset consisting of 276 real-world attack incidents and 91 ESM-related Code4rena findings. Based on these cases, we develop a cause-oriented taxonomy with four primary categories and eleven subcategories, capturing manipulated economic states and violated integrity assumptions.
    \item We systematically analyze the evaluation datasets used by six existing PM/ESM detection studies through our taxonomy. We find that existing benchmarks cover 6-10 of the 11 subcategories but remain concentrated on a limited set of vulnerability mechanisms.
    \item We evaluate seven representative detection approaches. The best-performing approach detects only 50.4\% of evaluated real-world attacks. Through taxonomy-guided FN and FP analysis, we identify the fundamental limitations of existing detection paradigms and derive practical recommendations and future directions for more generalizable ESM detection.
\end{itemize}

\section{Background and Motivation}

\subsection{Decentralized Finance}

Decentralized Finance (DeFi) offers a wide range of financial services through smart contracts. Among these applications, decentralized exchanges (DEXs) serve as fundamental components of the DeFi ecosystem by enabling permissionless cryptocurrency trading. 
DEXs facilitate token swaps through liquidity pools, where liquidity providers deposit pairs of tokens into smart contracts and users trade directly against these pooled reserves.
To determine asset prices within liquidity pools, DEXs typically rely on Automated Market Makers (AMMs), which define mathematical mechanisms for balancing token reserves and facilitating trades. The most widely adopted AMM model is the constant product market maker (CPMM), which maintains an invariant: $x \times y = k$, where $x$ and $y$ represent the reserves of two paired tokens in a liquidity pool, and k is a constant value. Under this mechanism, the relative price between two tokens is determined by the ratio of their reserves, making it sensitive to changes in pool reserves. 
Beyond decentralized exchanges, DeFi protocols provide various financial services, such as lending and staking platforms, which often rely on these economically derived values to perform critical operations. For example, lending protocols may use token prices obtained from DEXs or other oracle mechanisms to evaluate collateral values. Consequently, the correctness and reliability of these economic states are essential for the security of DeFi protocols.

\subsection{Motivating Example}
\label{sec:motivating-example}

\begin{figure}[htbp]
\centerline{\includegraphics[width=0.97\linewidth]{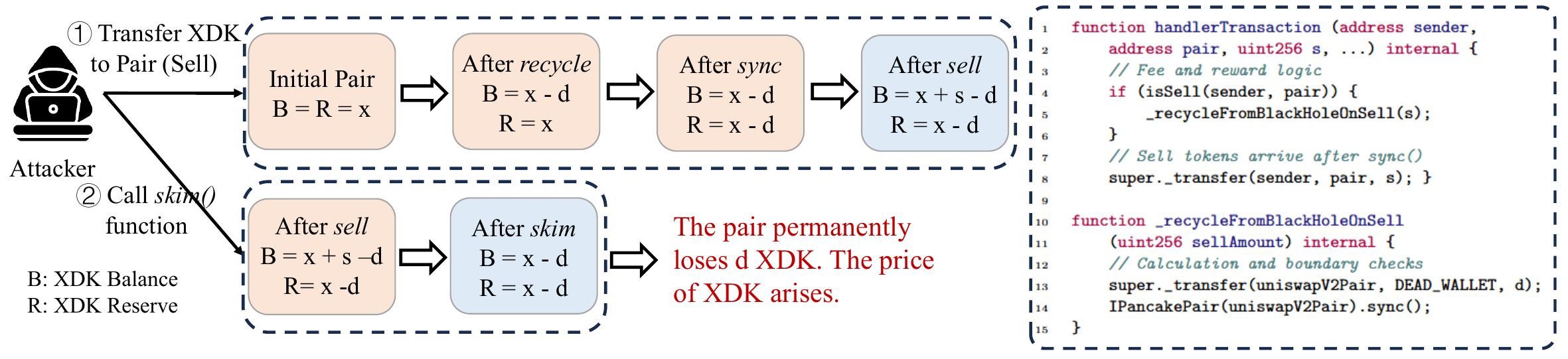}}
\caption{Motivating Example: XDK Incident}
\label{Motivating example}
\end{figure}

On February 16, 2026, XDK suffered an attack that exploited a vulnerability in its interaction with an XDK/GPC automated market maker pair~\cite{XDKIncident}. The XDK token implemented a \texttt{recycle} mechanism intended to process the liquidity associated with LP tokens permanently sent to a dead address. Conceptually, a correct recycling operation should preserve the consistency between LP ownership and the pool's underlying assets by redeeming the corresponding LP shares and proportionally withdrawing both XDK and GPC from the pool. 
However, XDK bypassed this procedure. When a user sold XDK into the pair, the \texttt{recycle} hook directly transferred XDK out of the live pool without burning LP tokens or proportionally withdrawing GPC. As shown in Fig.~\ref{Motivating example}, this operation created an economically invalid state transition between LP ownership and the pool reserves.

The attacker exploited this vulnerability through the interaction between \texttt{sync()} and \texttt{skim()}.
In a Uniswap V2-style AMM, \texttt{sync()} updates the records reserves to match the pool's actual token balances, while \texttt{skim()} transfers any excess balance beyond the recorded reserves.
Suppose the pair initially contains $x$ XDK, with its actual balance $B$ equal to the recorded reserve $R$. When the \texttt{recycle} hook removes $d$ XDK, the balance decreases to $B=x-d$ while the reserve remains $R=x$. The subsequent \texttt{sync()} call updates the reserve to $R=x-d$. 
Crucially, this occurs before the incoming transfer of $s$ XDK is completed. Once these tokens arrive, the balance becomes $B=x+s-d$, while the recorded reserve remains $R=x-d$. The attacker can then invoke \texttt{skim()} to recover the surplus $s$, restoring the balance to $B=R=x-d$. Thus, the pool loses $d$ XDK while the attacker recovers the tokens used to trigger the operation.

Specifically, the attacker first leveraged flashloan to borrow GPC and used it to buy XDK. Then, the attacker repeatedly executed this \texttt{transfer}--\texttt{recycle}--\texttt{sync}--\texttt{skim} cycle, progressively reducing the pool's XDK reserves without a corresponding reduction in GPC reserves. As the reserve ratio became increasingly distorted, the implied price of XDK relative to GPC rose. The attacker eventually exchanged the accumulated XDK for an inflated amount of GPC, repaid the flash loan, and retained the remaining assets as profit.


Although the attack eventually resulted in  a substantial distortion in XDK/GPC exchange price, the underlying vulnerability was not a manipulated oracle price or an incorrect price calculation. Instead, the \texttt{recycle} mechanism enabled unauthorized changes to the AMM's token balances, which were subsequently committed through \texttt{sync()} and exploited through \texttt{skim()}. The resulting price distortion was therefore a downstream consequence of violated economic-state integrity.
However, existing studies~\cite{liu2025llm,zhong2025detecting} often broadly categorize such incidents as \emph{price manipulation}, without distinguishing reserve-state manipulation from attacks that directly tamper with oracle prices or price-dependent computations. 
Such coarse-grained characterization can obscure the actual root causes and execution semantics of these attacks, potentially limiting the coverage of existing detection approaches. This case therefore highlights the importance of systematically revisiting and characterizing existing ``price manipulation'' attacks from a broader economic-state perspective, so as to better understand their underlying manipulation mechanisms and guide the design of more comprehensive and generalizable detection techniques.

\subsection{Definition of Economic-State Manipulation}
\label{sec:esm-definition}

We define an \emph{economic state} as a protocol-level value or relation whose integrity affects the economic outcome of a protocol operation. 
Economic states include, but are not limited to, oracle and market prices, accounting-derived exchange rates and valuations, AMM balances and reserves, LP-related states, and execution-time intermediate states.
Let $f$ be an economically sensitive operation, such as borrowing or liquidity withdrawal. For an execution trace $\tau$, we define the economic decision context of $f$ as:

\begin{equation}
\Gamma_f^\tau =
<
E_{f,\mathrm{pre}}^\tau,
G_f^\tau,
E_{f,\mathrm{post}}^\tau
>,
\end{equation}
where $E_{f,\mathrm{pre}}^\tau$ is the economic state observed or consumed by $f$, $G_f^\tau$ represents the economic constraints applicable to the execution, such as authorization conditions and price bounds, and $E_{f,\mathrm{post}}^\tau$ is the committed economic state resulting from the operation. Let $\Phi_f(\Gamma_f^\tau)$ denote the economic-integrity condition that the complete decision context of $f$ is expected to satisfy.

A protocol contains an \emph{Economic-State Manipulation (ESM) vulnerability} if there exist a feasible adversarial trace $\tau$ and an economically sensitive operation $f$ such that

\begin{equation}
\exists\;
\tau \in \mathcal{T}_{\mathrm{adv}},
\quad \mathrm{s.t.} \quad
\mathsf{Commit}_f(\tau)
\land
\neg \Phi_f(\Gamma_f^\tau)
\land
\Delta_f(\tau,\tau^\star) > \epsilon_f ,
\end{equation}
where $\mathcal{T}_{\mathrm{adv}}$ denotes the set of feasible traces through which an adversary can induce, amplify, or exploit an economic-integrity violation, and $\mathsf{Commit}_f(\tau)$ indicates that $f$ successfully commits an economically relevant state transition. $\tau^\star$ denotes an integrity-preserving counterfactual execution that performs the same intended operation under equivalent external conditions but without the integrity violation. $\Delta_f(\tau,\tau^\star)$ measures the protocol-specific economic deviation between the adversarial execution and its integrity-preserving counterpart, and $\epsilon_f$ is a context-specific materiality threshold.

An \emph{ESM attack} is a realized execution trace satisfying the above conditions. Intuitively, an ESM vulnerability exists when an adversary can induce or exploit a violation of an economic-integrity condition, an economically sensitive operation nevertheless accepts or commits under that condition, and the resulting outcome materially deviates from the intended economic outcome. 
To characterize the root cause of each ESM instance, we represent it using the tuple

\begin{equation}
\mathcal{I} = \langle V, A, S, M \rangle,
\end{equation}
where $V$ denotes the economically relevant value, relation, or state component involved in the violation; $A$ denotes the violated economic-integrity assumption captured by $\Phi_f$; $S$ denotes the economically sensitive sink that consumes or commits the affected state; and $M$ denotes the reusable causal mechanism through which the adversarial trace induces or exploits the violation.

This representation separates observable economic consequences from their underlying causes. Conventional price manipulation corresponds to the special case where $V$ is an explicit market price or price source. In broader ESM instances, however, $V$ may instead be an accounting rate, AMM reserve/balance state, liquidity state, or execution-time invariant. In such cases, any observed price distortion is only a downstream consequence of the underlying economic-state integrity violation.

\section{Methodology}

\begin{figure}[htbp]
\centerline{\includegraphics[width=0.99\linewidth]{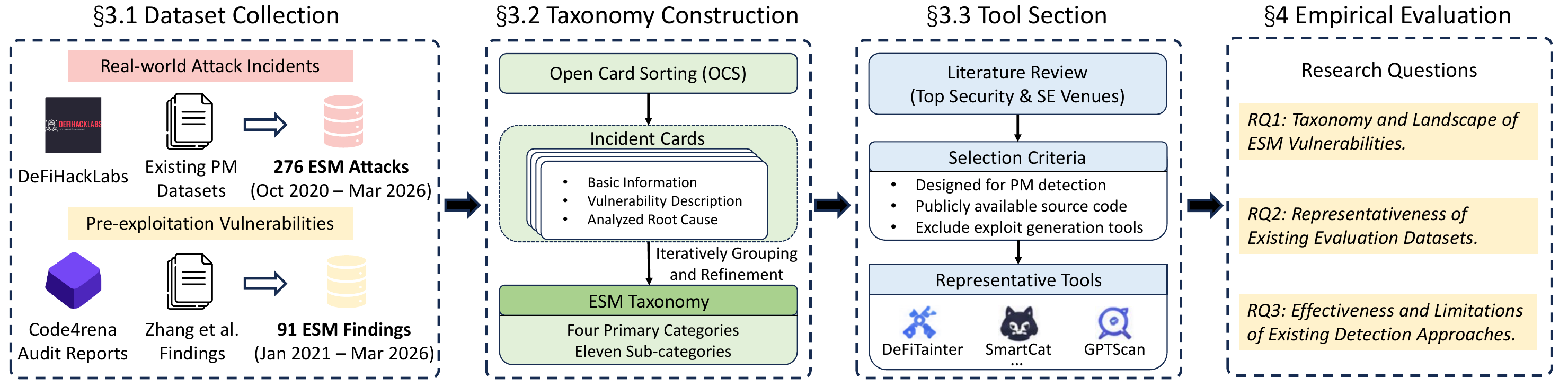}}
\caption{The Framework of Our Study}
\label{framework}
\end{figure}

Figure~\ref{framework} illustrates the overall framework of our study. First, we collect ESM vulnerabilities from real-world attack incidents and security audit reports. Second, we apply open card sorting to systematically analyze the collected cases and construct a cause-oriented taxonomy. Third, we conduct a literature review and apply predefined selection criteria to identify representative PM/ESM detection tools for evaluation. Finally, we conduct an empirical evaluation using the collected dataset, taxonomy, and selected approaches to answer our research questions.

\subsection{Dataset Collection}

To systematically investigate ESM attacks in DeFi, we conduct a dataset from two complementary sources: (1) real-world exploit incidents collected from DeFiHackLabs and existing price manipulation datasets, and (2) vulnerabilities identified before exploitation in Code4rena audit reports.. Combining these two sources enables us to capture both attacks that have been successfully exploited in the wild and latent vulnerabilities identified before exploitation, thereby providing a broader and more representative view of ESM mechanisms in practice.

We first collected all 754 publicly documented security incidents in DeFiHackLabs up to March 10, 2026. DeFiHackLabs maintains a comprehensive repository of real-world blockchain securiy incidents, providing attack descriptions and executable proof-of-concept (PoC) scripts. We first excluded incidents whose root causes were clearly attributable to other well-established vulnerability classes, such as reentrancy and access control vulnerability. 
For the remaining candidates, two researchers independently reviewed the incident descriptions and PoCs, and further inspected the corresponding transaction traces using Phalcon~\cite{Phalcon} when necessary. Disagreements were resolved through discussion. In addition, we incorporated real-world PM incidents reported in existing datasets published at top-tier security and software engineering venues detailed in Section~\ref{tool}. After deduplication and manual validation, this process resulted in 276 real-world ESM incidents spanning Oct 2020 to Mar 2026, with an aggregated loss of approximately \$524 million.

To complement exploited incidents with vulnerabilities discovered before exploitation, we further collected high-severity findings from Code4rena, a widely used competitive smart-contract auditing platform. For audit reports published before 2023, we incorporated the findings curated by Zhang et al.~\cite{zhang2023demystifying}. We then extended the collection by systematically reviewing Code4rena reported from January 2023 to March 2026. In total, we examined 1,283 high-severity audit findings. Two authors independently inspected each finding based on its vulnerability description and determined whether it satisfied our ESM definition. After resolving disagreements through discussion, we obtained 91 ESM-related audit findings. Together with the real-world attack incidents, these findings provide complementary observations of both exploited and proactively disclosed ESM vulnerabilities. 

\subsection{Taxonomy Construction}

To systematically characterize the root causes of ESM vulnerabilities, we construct a taxonomy based on the underlying manipulation mechanisms. Existing smart-contract vulnerability taxonomies are largely designed around general vulnerability classes and provide limited characterization of manipulation vulnerabilities~\cite{zhang2023demystifying,liu2025have, chaliasos2024smart, cernera2023token}. The most closely related study, DeFiRanger~\cite{wu2023defiranger}, analyzed the root causes of 41 price manipulation attacks collected before March 2023. However, its relatively small and early-stage dataset limits both the diversity and timeliness of the identified attack patterns. This motivates a more comprehensive taxonomy to characterize ESM vulnerabilities and support systematic evaluation of existing detection approaches.

Consequently, we analyze the collected dataset to construct our taxonomy using \emph{open card sorting (OCS)}, a qualitative method widely used to identify recurring patterns and organize observations based on shared characteristics~\cite{fincher2005making,zimmermann2016card}. Specifically, we create an individual card for each attack or vulnerability finding. Each card contains three types of information: (1) \emph{basic incident information}, including the affected protocol, incident date, vulnerable contract, and attack consequence, (2) \emph{report-derived vulnerability description}, summarizing the vulnerable behavior and exploitation process based on the original incident analysis, PoC, or audit report; and (3)\emph{root-cause analysis}, identifying the affected economic state, the violated integrity assumption, and the underlying manipulation mechanism according to our ESM definition. This representation enables us to compare vulnerabilities based on their underlying cases rather than their reported labels or observable consequences.

The sorting process involved three researchers. First, two researchers, each with two years of experience in blockchain security, independently reviewed the cards and grouped them according to their underlying root causes. During this stage, they assigned descriptive labels to each group. The two researchers then jointly compared their initial classifications and iteratively refined the taxonomy without relying on predefined categories. To avoid introducing overly specific or weakly supported categories, any newly proposed category had to be reviewed and confirmed by a third researcher with extensive industry experience in blockchain security. The third researcher also served as the adjudicator for unresolved disagreements between the first two researchers.
Following this procedure, we developed a cause-oriented taxonomy consisting of four primary categories and eleven subcategories of ESM vulnerabilities detailed in Section~\ref{taxonomy}.

\subsection{Tool Selection}\label{tool}

To collect representative manipulation vulnerability detection approaches, we conduct a literature review of academic publications from top-tier software engineering and security venues (e.g., ICSE, ASE, FSE, ISSTA, Usenix Security, and TDSC). We focus on approaches specifically designed to detect PM vulnerabilities or provide analysis capabilities applicable to PM detection. We apply two criteria to select candidate tools. First, the tool must be explicitly designed to detect PM vulnerabilities or originally proposed to analyze PM-related security issues. We exclude approaches that focus on automatically generating PM exploits or synthesizing PM attacks (e.g., FlashSyn~\cite{chen2024flashsyn}). Second, the tool must provide publicly available source code, enabling reproducible and large-scale evaluation on our dataset. We further exclude general-purpose smart contract analysis tools without PM-specific detection capabilities (e.g., Slither~\cite{feist2019slither}, Oyente~\cite{luu2016making}, and Mythril~\cite{sharma2022survey}). 

Based on these criteria, we identify seven representative
PM detection tools, which fall into three categories according to their analysis inputs and detection mechanisms.

\begin{itemize}
    \item \emph{Transaction-dependent analysis:} DeFiRanger~\cite{wu2023defiranger}, DeFort~\cite{xie2024defort}, and DeFiScope~\cite{zhong2025detecting} rely on transaction-level behaviors to identify manipulation attacks. DeFiRanger reconstructs cash-flow trees from transaction traces and matches them against predefined attack patterns. DeFort detects manipulation based on observable signals such as abnormal price changes and attack profits, and DeFiScope reconstructs DeFi operations from transaction traces and identifies suspicious manipulation behaviors.
    \item \emph{Bytecode analysis:} DeFiTainter~\cite{kong2023defitainter} and SmartCat~\cite{zhang2025following} analyze contract bytecode to identify potential manipulation vulnerabilities. DeFiTainter employs static taint analysis to track potentially manipulated values from predefined sources to security-critical sinks, whereas SmartCat performs inter-contract analysis to identify attacker-controlled contracts and potential manipulation behaviors.
    \item \emph{LLM-assisted sourcecode analysis:} GPTScan~\cite{sun2024gptscan} and PMDetector~\cite{liu2025llm} combine static analysis with large language models to assess potential vulnerabilities. GPTScan employs rule-guided candidate selection followed by LLM-based semantic analysis, while PMDetector constructs source-to-sink taint paths and uses an LLM to assess their exploitability. 
\end{itemize}


\section{Evaluation Results}

\subsection{RQ1:Taxonomy and Landscape of ESM Vulnerabilities}\label{taxonomy}

To answer RQ1, we analyze the collected ESM dataset and develop a root-cause-oriented taxonomy of ESM vulnerabilities. Unlike existing classifications that mainly categorize vulnerabilities based on their observable consequences (e.g., abnormal price fluctuations) or attack techniques (e.g., flash loans), our taxonomy focuses on the underlying economic-integrity assumptions violated by these vulnerabilities. This perspective enables us to distinguish the manipulated economic states from the mechanisms used to manipulate them and the consequences observed during exploitation.

As shown in Table~\ref{tab:taxonomy-formal}, our taxonomy comprises four primary categories: \textit{Price-Source Trust}, \textit{Protocol-Internal Accounting}, \textit{AMM and Liquidity State Integrity}, and \textit{Execution-Time Protection}. These categories are further divided into eleven subcategories, capturing the diverse root causes and manipulation mechanisms observed in 367 real-world ESM cases. For each subcategory, we characterize the economically relevant state ($V$) and the corresponding integrity assumption ($A$) that the protocol is expected to uphold. This characterization provides a consistent basis for analyzing how different ESM vulnerabilities arises and how their underlying causes differ.

\begin{table*}[t]
\centering
\footnotesize
\caption{Root-Cause Taxonomy of ESM Vulnerabilities and Their Economic-Integrity Assumption}
\label{tab:taxonomy-formal}

\resizebox{\textwidth}{!}{
\begin{tabular}{
>{\raggedright\arraybackslash}p{2.2cm}
>{\raggedright\arraybackslash}p{3.3cm}
>{\raggedright\arraybackslash}p{3.2cm}
>{\raggedright\arraybackslash}p{4.5cm}
}

\toprule
\textbf{Category} & \textbf{Type} & \textbf{V: economic state} & \textbf{A: violated assumption} \\
\midrule

\multirow{6}{=}{Price-Source Trust}
& A1: Unsafe Direct Market-State Pricing
& Live market-state price & V reflects the asset's fair exogenous value\\

& A2: Oracle Construction or Validation Defect
& Oracled-reported price or feed & The pipeline producing V is correctly composed, normalized, validated \\

& A3: Static or Stale Reference Dependence
& Hardcoded or long-lived reference & V is reconciled with the live market before use \\

\midrule

\multirow{6}{=}{Protocol-Internal Accounting}
& B1: Claim-to-Backing Conversion Violation
& Claim-to-Backing conversion rate & Minting or redemption preserve proportional backing \\

& B2: Composite or Position Misvaluation
& Value of a composite asset or position & Aggregation faithfully reflects the economic value of its components \\

& B3: Entitlement or Obligation Misattribution
& User-specific reward, fee or claim & V equals the amount attributable to the user under intended rules \\

\midrule

\multirow{6}{=}{AMM and Liquidity State Integrity}
& C1: Unsafe Token-side AMM State Mutation
& AMM pair token balances & Balances change only through intended AMM operations \\

& C2: LP-State Integrity Violation
& Pool invariant and LP-supply relationship & Liquidity operations preserve the invariant and the LP-state relationship \\

& C3: Unsafe Balance-Reserve Settlement
& Balance-reserve correspondence & Recorded reserves remain consistent with actual balances \\

\midrule

\multirow{4}{=}{Execution-Time Protection}
& D1: Ineffective Execution Output Constraint
& The output bound in $G_f$ & $G_f$ contains an enforceable, independent bound \\

& D2: Premature Consumption of Unsettled State
& State consumed before settlement into $E_{f,\mathrm{post}}$ & State is settled before any security-critical read \\

\bottomrule
\end{tabular}
}
\end{table*}

\subsubsection{Definitions of ESM Vulnerability types}
This section elaborates on the four primary categories and their subcategories, describing their underlying root causes, vulnerability mechanisms, and representative examples. Table~\ref{tab:taxonomy-formal} summarizes the affected economic states and violated integrity assumptions for each vulnerability type.

\noindent\textbf{Price-Source Trust.}
This category captures vulnerabilities where a protocol relies on price-related values as trusted economic references, while the underlying price source fail to satisfy the required integrity guarantees. In these attacks, the manipulated object is typically an explicit value or price reference consumed by security-critical operations, such as lending, liquidation, or asset exchange. The fundamental assumption violated in this category is that the consumed price is reliable, fresh, and resistant to manipulation. However, attackers may manipulate the underlying price source or exploit weakness in its construction and validation process, causing the protocol to make incorrect economic decisions. We further classify these vulnerabilities into three subcategories.

(1) Unsafe Direct Market-State Pricing. This subcategory captures vulnerabilities where a protocol directly use manipulable market states, such as AMM reserve ratios or spot quotes, as authoritative price references. 
The violated assumption is that the observed market-derived values reliably reflect asset prices, even although attackers can temporarily distort them before price-sensitive operations.

(2) Oracle Construction or Validation Defect. 
This subcategory covers vulnerabilities arising from flaws in the construction of validation of oracle-derived prices. Unlike RC-A1, which directly consumes market states, these vulnerabilities involve derived price or exchange rates whose integrity is compromised by incorrect source selection, data aggregation, or validation. Such weaknesses in the oracle pipeline or its consumer-side checks allow attackers to influence the prices consumed by downstream operations.

(3) Static or Stale Reference Dependence.
This category captures vulnerabilities where a protocol relies on a fixed or insufficiently updated price references that no longer reflect current market conditions. The underlying weakness does not necessarily involve active manipulation of a live price source. Instead, attackers exploit discrepancies between outdated references and current market values to influence price-sensitive operations.

Example (RC-A1): as shown in Fig.~\ref{lst:taxonomy-code-examples} (a), the EGD Finance attacker used flashloan funed trades to manipulate the reserves of the EGD/USDT liquidity pool.
The protocol's reward mechanism directly derived the EGD price directly from the manipulated reserve ratio, resulting in excessive reward distribution. Thus, the root cause was the use of manipulable market states as authoritative price references, while the reward calculation served as the economically sensitive operations affected by the manipulation.

\definecolor{TaxBlue}{HTML}{D9E8FF}
\definecolor{TaxGreen}{HTML}{E1F3E6}
\definecolor{TaxOrange}{HTML}{FFE9CF}
\definecolor{TaxRed}{HTML}{FFE0E0}
\definecolor{TaxGray}{HTML}{F2F2F2}
\definecolor{TaxInk}{HTML}{1D1D1F}
\definecolor{TaxCodeBg}{HTML}{F8F8F8}
\definecolor{TaxCodeKw}{HTML}{005CC5}
\definecolor{TaxCodeCm}{HTML}{6A737D}

\lstdefinestyle{taxonomySolidity}{
  language=Java,
  basicstyle=\ttfamily\fontsize{7pt}{7.5pt}\selectfont,
  keywordstyle=\color{TaxCodeKw}\bfseries,
  commentstyle=\color{TaxCodeCm},
  stringstyle=\color{red!50!black},
  numbers=left,
  numberstyle=\tiny\color{TaxCodeCm},
  numbersep=6pt,
  frame=lines,
  framerule=0.4pt,
  rulecolor=\color{black!55},
  backgroundcolor=\color{TaxCodeBg},
  showstringspaces=false,
  columns=fullflexible,
  keepspaces=true,
  breaklines=true,
  tabsize=2,
  xleftmargin=1.8em,
  framexleftmargin=1.6em
}

\begin{figure*}[htbp]
\centering
\begin{minipage}[t]{0.485\textwidth}
{\centering\textbf{(a) EGD Finance Incident (RC-A1)}\par}
\begin{lstlisting}[style=taxonomySolidity]
IPair public pair;        // EGD/USDT pair
IERC20 public egd;
mapping(address => uint256) public quota;

function claimReward() external {
    // attacker-manipulated AMM reserves
    (uint112 rEGD, uint112 rUSDT,) =
        pair.getReserves();

    // mutable state is converted to a price
    uint256 price =
        uint256(rUSDT) * 1e18 / uint256(rEGD);

    // reward amount is inflated
    uint256 reward =
        quota[msg.sender] * 1e18 / price;
    egd.transfer(msg.sender, reward);
}
\end{lstlisting}
\end{minipage}
\hfill
\begin{minipage}[t]{0.485\textwidth}
{\centering\textbf{(b) Onyx Protocol Incident (RC-B1)}\par}
\begin{lstlisting}[style=taxonomySolidity]
function exchangeRate()
    public view returns (uint256) {   
    // direct donations increase cash
    uint256 cash =
        underlying.balanceOf(address(this));

    // tiny supply amplifies claim value
    return (cash + totalBorrows
        - totalReserves) * 1e18
        / totalSupply;
}
function collateralValue(address user)
    public view returns (uint256)
{
    // inflated rate enters collateral valuation
    return oTokenBalance[user]
        * exchangeRate() * oraclePrice / 1e36;
}
\end{lstlisting}
\end{minipage}
\caption{Representative Examples of Category RC-A1 and RC-B1}
\label{lst:taxonomy-code-examples}
\end{figure*}

\noindent\textbf{Protocol-Internal Accounting.} This category captures vulnerabilities arising from incorrect economic relationships within a protocol's internal accounting logic.
Unlike \textit{Price-Source Trust}, these vulnerabilities do not fundamentally depend on unreliable external price inputs. Even when external prices are accurate, flaws in protocols can produce economically inconsistent outcomes. The underlying root cause is therefore a violation of protocol-specific accounting invariants rather then the consumption of manipulated external prices. 
We further divide this category into three subcategories.

(1) Claim-to-Backing Conversion Violation. 
This subcategory captures vulnerabilities where the conversion relationship between protocol-issued claims and their underlying backing assets becomes economically inconsistent.
Typical claims include vault shares, staking claims, and other redeemable representations of underlying assets. Such vulnerabilities arise when minting, redemption, or backing calculations fail to preserve the intended proportional relationships between outstanding claims and their recognized assets.

(2) Composite or Position Misvaluation. 
This subcategory covers vulnerabilities where a protocol incorrectly values composite assets or financial positions, even when their individual component prices are accurate. 
Typical causes include omitted or double-counted components, incorrect units, or improper weighting in valuation formulas. 
These errors can affect LP positions, asset portfolios, and collateral-debt positions, resulting in valuation that do not accurately reflect their underlying economic composition.

(3) Entitlement or Obligation Misattribution. This subcategory captures cases where a protocol incorrectly assigns economic credits or obligations to a specific user or account. The aggregate economic value may be correct, but the attribution of the value violates the intended ownership, profit realization or debt accounting rules. Such inconsistencies may allow attackers to claim excessive benefits or avoid legitimate obligations.

Example (RC-B1): as shown in Fig.~\ref{lst:taxonomy-code-examples} (b), the Onyx protocol calculated the \emph{oToken} exchange rate using the ratio between the market's underlying token balance and the token supply. The attacker directly transferred underlying tokens into the market without minting additional \emph{oTokens}, increasing the asset balance used in the exchange-rate calculation. With a small outstanding oToken supply, this balance increase substantially inflated the exchange rate.
The protocol then used this manipulated conversion rate to value the collateral, enabling excessive borrowing. 

\noindent\textbf{AMM and Liquidity State Integrity.} 
This category captures ESM vulnerabilities arising from violations of the consistency or authorization constraints governing AMM token balances, recorded reserves, and liquidity ownership.
Unlike conventional PM attacks, the underlying vulnerability does not necessarily involve an unreliable price source. Instead, it arises when token or liquidity operations introduce economically invalid state transitions that violate the intended AMM accounting model. Such violations may subsequently distort exchange prices or enable unauthorized asset extraction.
We divide this category into three subcategories.

(1) Unsafe Token-Side AMM State Mutation. This subcategory captures vulnerabilities where token-side behavior or authorization-mediated operations modify AMM token balances outside the intended accounting and authority model.
Representative mechanisms include direct balance manipulation, fee-on-transfer logic, or public functions capable of changing pair balances. The violated assumption is that token-side operations cannot induce unauthorized changes to the pool's economic state that bypass its intended swap and liquidity mechanisms.

(2) LP-State Integrity Violation. The subcategory refers to vulnerabilities where liquidity-related operations violate the intended relationship between LP token supply, liquidity ownership, and underlying pool assets. 
Representative patterns include incorrect add/remove liquidity logic, double accounting of pool assets, and unauthorized reduction of pool-side assets. Such defects can cause minted LP shares or redeemed assets to deviate from the liquidity legitimately contributed or owned.

\begin{figure*}[htbp]
\centering
\begin{minipage}[t]{0.485\textwidth}
\vspace{0pt}
{\centering
\textbf{(a) Caviar H-02 Finding (RC-C2)}
\par}
\begin{lstlisting}[
  style=taxonomySolidity
]
function add(uint256 baseAmt, uint256 fracAmt)
    public returns (uint256 lpAmt)
{
    uint256 supply = lpToken.totalSupply();

    uint256 baseShare =
        baseAmt * supply / baseReserves();
    uint256 fracShare =
        fracAmt * supply / fracReserves();

    // LP issuance uses only the smaller side
    lpAmt = Math.min(baseShare, fracShare);

    // both full amounts are still accepted
    _transferFrom(msg.sender,
        address(this), fracAmt);
    ERC20(baseToken).safeTransferFrom(
        msg.sender, address(this), baseAmt);
    lpToken.mint(msg.sender, lpAmt);
}
\end{lstlisting}
\end{minipage}
\hfill
\begin{minipage}[t]{0.485\textwidth}
\vspace{0pt}
{\centering
\textbf{(b) Kashi Borrow Incident (RC-D2)}
\par}
\begin{lstlisting}[
  style=taxonomySolidity
]
function borrow(address to, uint256 amount)
external
{
    accrueInterest();

    // record the new debt
    debt[msg.sender] += amount;

    // BUG: solvency is checked using the
    // previously cached exchange rate
    require(
        isSolvent(msg.sender, exchangeRate),
        "insolvent"
    );
    asset.transfer(to, amount);

    // price-sensitive state is refreshed
    // only after the loan is committed
    exchangeRate = oracle.getRate();
}
\end{lstlisting}
\end{minipage}

\caption{Representative Examples of Category RC-C2 and RC-D2}
\label{fig:taxonomy-b1-c2-examples}
\end{figure*}

(3) Unsafe Balance-Reserve Settlement. This subcategory captures cases where attackers exploit inconsistencies between actual AMM token balances and recorded reserves through unsafe synchronization or settlement mechanisms.
Although temporary balance-reserve differences can occur during normal AMM operations, vulnerabilities arise when these discrepancies are incorrectly settled or exploited to produce unintended economic outcomes. Representative mechanisms include unsafe interactions involving \emph{sync} or \emph{skim}, as well as improper reserve updates that allow attackers to convert temporary balance inconsistencies into economic gains. 

Example (RC-C2): in the Caviar H-02 Finding (Fig.~\ref{fig:taxonomy-b1-c2-examples} (a)), the protocol independently calculated the LP-token shares corresponding to the two deposited assets and minted LP tokens based on the smaller proportional contribution. However, the \texttt{add} function transferred the full amounts of both assets into the pool without accounting for the excess contribution. When the deposited amounts deviated from the current reserve ratio, the excess portion of the over-supplied side increased the pool reserves without granting the provider corresponding LP claim.
Consequently, the economic value of the excess contribution effectively accrued to existing LP holders.

\noindent\textbf{Execution-Time Protection.} 
This category captures ESM vulnerabilities where a protocol commits a sensitive operation without enforcing the safeguards required at execution time.
The vulnerability arises from either an ineffective constraint on the realized output or unsafe ordering between state refresh, settlement, and consumption.
The violated assumption is that an economic operation should only commit when its outcomes is independently bounded and all state used in the decision has been properly refreshed and settled. We divide this category into two subcategories.

(1) Ineffective Execution Output Constraint. This subcategory captures vulnerabilities where protocol-controlled economic operations execute without effective constraints on its realized output. Representative operations include swaps and liquidity operations. Typical failure patterns include missing or zero \emph{minOut} values and attacker-controlled slippage parameters.

(2) Premature Consumption of Unsettled State. This subcategory captures vulnerabilities where sensitive-critical operations consume economic states before the required refresh, update, or settlement has been completed. Typical patterns include check-before-refresh, read-before-update, and unsafe ordering between state updates and external interactions. Such defects allow protocol components to perform valuation, solvency checks, or accounting based on intermediate states that do not reflect the conditions required for a valid economic decision.

Example (RC-D2): in the Kashi incident (Fig.~\ref{fig:taxonomy-b1-c2-examples} (b)), the protocol involved a borrowing operation that performed a solvency check using an outdated exchange data. 
Specifically, the borrow function updated the user's debt position and evaluated solvency using a cached exchange rate before refreshing the price-sensitive state. Consequently, the protocol could approve a borrowing operation that would fail the solvency check under the updated exchange rate, allowing the borrower to obtain assets beyond the intended collateral constraints. The root cause was therefore the premature consumption of an unsettled economic state during a security-critical decision. 

\begin{resultbox}
\textbf{Finding 1: }
Our analysis categorizes 276 real-world attacks and 91 audit findings into four primary root-cause categories of ESM vulnerabilities: \textit{Price-Source Trust}, \textit{Protocol-Internal Accounting}, \textit{AMM and Liquidity State Integrity}, and \textit{Execution-Time Protection}, further divided into eleven subcategories.
\end{resultbox}

\begin{figure}[htbp]
    \centering
    \begin{subfigure}{0.46\linewidth}
        \centering
        \includegraphics[width=\linewidth]{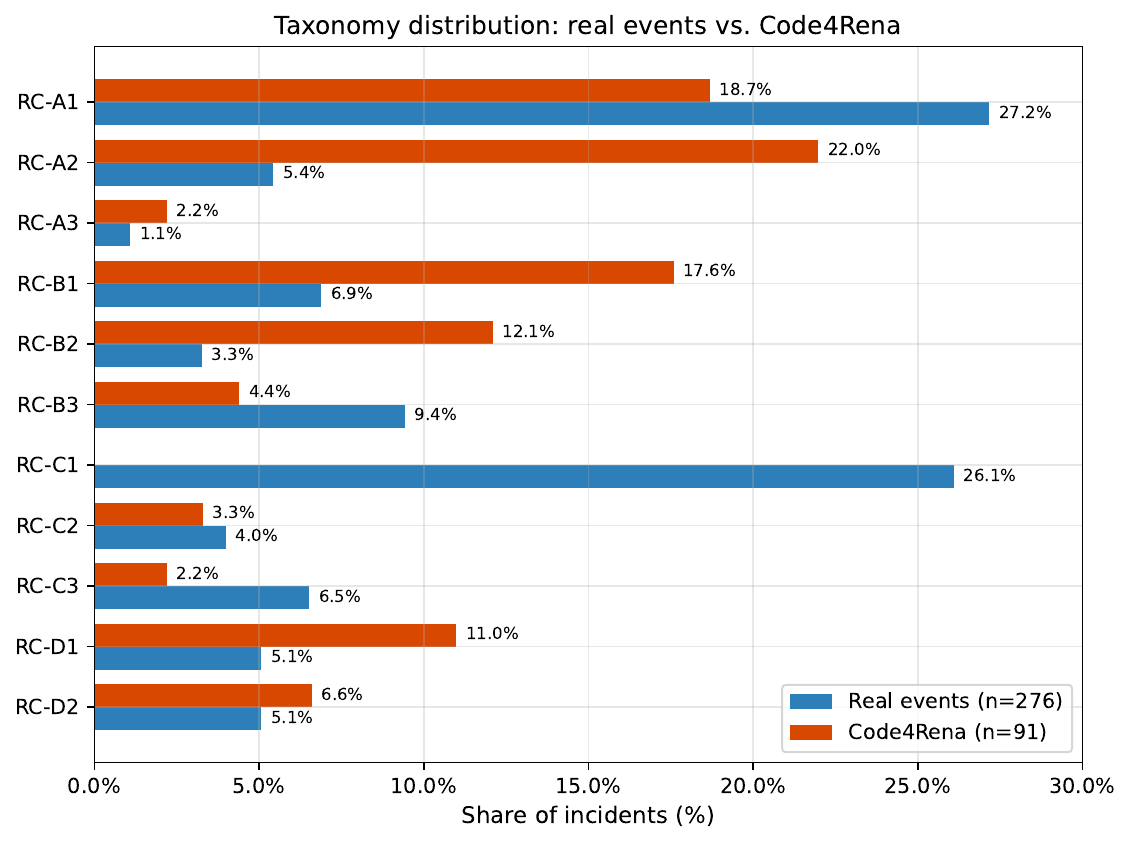}
        \caption{Root-cause distribution across datasets}
        \label{fig:root_distribution}
    \end{subfigure}
    \hfill
    \begin{subfigure}{0.46\linewidth}
        \centering
        \includegraphics[width=\linewidth]{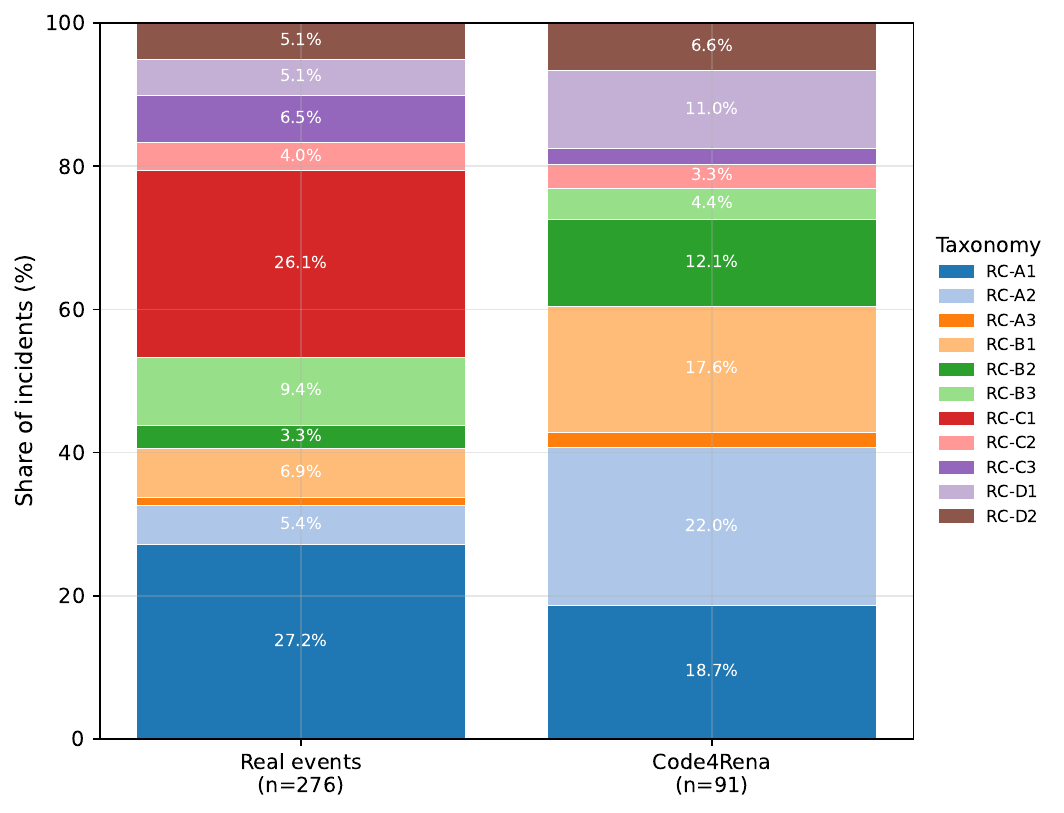}
        \caption{Root-cause composition across datasets}
        \label{fig:subcategory_distribution}
    \end{subfigure}
    \caption{Distribution and Composition of ESM Vulnerabilities.}
    \label{fig:esm_distribution}
    \vspace{-0.2cm}
\end{figure}

\subsubsection{Distribution of ESM Vulnerabilities}
To better understand the prevalence of different ESM vulnerabilities in practice, we analyze the distribution of different vulnerability categories. Specifically, we investigate the distribution of ESM vulnerabilities across different root-cause categories and subcategories. Fig.~\ref{fig:root_distribution} presents the distribution of individual subcategories, while Fig.~\ref{fig:subcategory_distribution} illustrates their overall composition in the two datasets.

As shown in Fig.~\ref{fig:root_distribution}, ESM vulnerabilities exhibit substantial diversity across different root causes, extending conventional price manipulation. Attackers can exploit a broad range of violated economic assumptions, including inconsistent protocol accounting, corrupted AMM states, and unsafe execution states. Among real-world attacks, \emph{Price-Source Trust} and \emph{AMM and Liquidity State Integrity} represent the most prevalent categories. Specifically, RC-A1 (\emph{Unsafe Direct Market-State Pricing}) accounts for 27.2\% of incidents, followed closely by RC-C1 (\emph{Unsafe Token-Side AMM State Mutation}) at 26.1\%. Together, these two subcategories account for more than half of the observed attacks. Beyond these dominant types, other root causes also contribute substantially to real-world exploits.
For example, RC-B3 (\emph{Entitlement or Obligation Misattribution}) accounts for 9.4\% of incidents, while RC-D1 and RC-D2 (\emph{Execution-Time Protection}) collectively account for 10.2\%. These results demonstrate that ESM vulnerabilities involve diverse violations of economic-integrity assumptions, with price distortion often arising as a downstream consequence rather than the underlying cause.


We further compare the root-cause distributions of real-world exploits and Code4rena audit findings. As shown in Fig.~\ref{fig:subcategory_distribution}, the two datasets exhibit substantially distinct vulnerability compositions, reflecting the different types of economic-integrity violations identified through exploitation and proactive security assessments. 
For real-world incidents, attackers tend to prioritize vulnerabilities involving economic states that can be directly manipulated and converted into immediate financial gains, particularly market-derived valuations and AMM states. In contrast, Code4Rena findings are less dominated by a few attack patterns and contain logic-related vulnerabilities. Notably, \emph{Unsafe Token-side AMM State Mutation} (RC-C1), which represents one of the most common causes in real-world exploits, is rarely observed in Code4rena findings. This difference suggests that security-conscious projects may have developed stronger awareness and defensive mechanisms against explicit token-level manipulation patterns. However, Code4rena findings contain a substantially higher proportion of protocol-internal accounting failures, indicating that complex economic logic remains a challenging source of vulnerabilities. For example, \emph{Protocol-Internal Accounting} vulnerabilities (RC-B) account for 34.1\% of audit findings.
Such accounting-related vulnerabilities are less likely to be immediately discovered and exploited in real-world attacks.

\begin{resultbox}
    \textbf{Finding 2:} ESM vulnerabilities extend far conventional price manipulation and arise from diverse economic-integrity violations. While real-world exploits are concentrated in market-derived price and AMM state manipulation, audit findings reveals a greater prevalence of oracle-related and protocol-internal accounting weaknesses, highlighting the diversity of underlying ESM root causes.
\end{resultbox}

\subsection{RQ2: Taxonomy Coverage of Available Evaluation Datasets}
To understand whether existing evaluation datasets can faithfully represent the diversity of ESM vulnerabilities observed in practice, we analyze the datasets used by the detection approaches identified in Section~\ref{tool}. We exclude GPTScan because it does not provide a dedicated PM evaluation dataset. Specifically, we map each real-world vulnerability incidents contained in existing dataset to our proposed ESM taxonomy and examine both category coverage and distributional balance. 
Table~\ref{tab:esm-taxonomy-distribution} summarizes the taxonomy distribution of existing evaluation datasets. 

\begin{table*}[t]
  \centering
  \caption{Distribution of Incidents across the ESM Taxonomy.}
  \label{tab:esm-taxonomy-distribution}
  \scriptsize
  \setlength{\tabcolsep}{2.5pt}
  \renewcommand{\arraystretch}{1.15}
  \resizebox{\textwidth}{!}{%
    \begin{tabular}{lcccccccccccc}
      \toprule
      \multirow{2}{*}{Dataset}
        & \multicolumn{3}{c}{RC-A}
        & \multicolumn{3}{c}{RC-B}
        & \multicolumn{3}{c}{RC-C}
        & \multicolumn{2}{c}{RC-D}
        & \multirow{2}{*}{Total} \\
      \cmidrule(lr){2-4}
      \cmidrule(lr){5-7}
      \cmidrule(lr){8-10}
      \cmidrule(lr){11-12}
        & RC-A1 & RC-A2 & RC-A3
        & RC-B1 & RC-B2 & RC-B3
        & RC-C1 & RC-C2 & RC-C3
        & RC-D1 & RC-D2 & \\
      \midrule

      DeFiRanger
        & 31.7\% (13)
        & 2.4\% (1)
        & 0.0\% (0)
        & 7.3\% (3)
        & 2.4\% (1)
        & 0.0\% (0)
        & 48.8\% (20)
        & 0.0\% (0)
        & 2.4\% (1)
        & 4.9\% (2)
        & 0.0\% (0)
        & 41 \\

      DeFiTainter
        & 47.8\% (11)
        & 0.0\% (0)
        & 0.0\% (0)
        & 8.7\% (2)
        & 4.3\% (1)
        & 21.7\% (5)
        & 13.0\% (3)
        & 0.0\% (0)
        & 4.3\% (1)
        & 0.0\% (0)
        & 0.0\% (0)
        & 23 \\

      DeFiScope
        & 29.5\% (28)
        & 4.2\% (4)
        & 0.0\% (0)
        & 5.3\% (5)
        & 4.2\% (4)
        & 7.4\% (7)
        & 30.5\% (29)
        & 3.2\% (3)
        & 7.4\% (7)
        & 5.3\% (5)
        & 3.2\% (3)
        & 95 \\

      DeFort
        & 31.5\% (17)
        & 11.1\% (6)
        & 0.0\% (0)
        & 3.7\% (2)
        & 3.7\% (2)
        & 11.1\% (6)
        & 31.5\% (17)
        & 0.0\% (0)
        & 3.7\% (2)
        & 3.7\% (2)
        & 0.0\% (0)
        & 54 \\

      SmartCat
        & 35.7\% (30)
        & 2.4\% (2)
        & 0.0\% (0)
        & 6.0\% (5)
        & 2.4\% (2)
        & 6.0\% (5)
        & 28.6\% (24)
        & 4.8\% (4)
        & 7.1\% (6)
        & 2.4\% (2)
        & 4.8\% (4)
        & 84 \\

    PMDetector
        & 35.6\% (26)
        & 6.9\% (5)
        & 0.0\% (0)
        & 15.1\% (11)
        & 2.7\% (2)
        & 9.6\% (7)
        & 12.3\% (9)
        & 5.5\% (4)
        & 5.5\% (4)
        & 1.4\% (1)
        & 5.5\% (4)
        & 73 \\

      \textbf{Ours}
        & \textbf{25.1\% (92)}
        & \textbf{9.5\% (35)}
        & \textbf{1.4\% (5)}
        & \textbf{9.5\% (35)}
        & \textbf{5.4\% (20)}
        & \textbf{8.2\% (30)}
        & \textbf{19.6\% (72)}
        & \textbf{3.8\% (14)}
        & \textbf{5.4\% (20)}
        & \textbf{6.5\% (24)}
        & \textbf{5.4\% (20)}
        & \textbf{367} \\

      \bottomrule
    \end{tabular}%
  }
\end{table*}

Our analysis shows that the examined datasets cover between 6 and 10 of the 11 identified subcategories. DeFiTainter, one of the earliest studies targeting price manipulation vulnerabilities, exhibits the narrowest taxonomy representation, covering only six categories. DeFiRanger and DeFort cover seven and eight categories, respectively. More recent datasets, including those used by DeFiScope, SmartCat, and PMDetector, achieve broader category-level coverage by including instances from ten subcategories. However, RC-A3 (\emph{Static or Stale Reference Dependence}) is absent from all existing datasets. These results indicate that existing benchmarks capture a relatively broad range of ESM root causes, although certain vulnerability types remain underrepresented.

Although existing datasets span multiple taxonomy categories, we observe substantial imbalances in the distribution of vulnerability types. In particular, RC-A1 (\emph{Unsafe Direct Market-State Pricing}) and RC-C1 (\emph{Unsafe Token-Side AMM State Mutation}) 
account for a large proportion of existing benchmark cases. Together, these two subcategories represent 80.5\% of DeFiRanger, 60.9\% of DeFiTainter, 60.0\% of DeFiScope, 63.0\% of DeFort, 64.3\% of SmartCat, and 47.9\% of PMDetector. Such concentration indicates that broad category coverage does not necessarily imply balanced representation of different ESM mechanisms.

Importantly, vulnerabilities within the taxonomy may exhibit similar observable consequences while arising from fundamentally different economic assumptions and exploitation mechanisms. For example, RC-A1 involves the use of manipulable market-derived prices as authoritative references, whereas RC-C1 represents a distinct class of vulnerabilities caused by token-side operations that violate the intended AMM state-transition constraints. 
These vulnerabilities differ in manipulated economic states and violated integrity assumptions. Consequently, the presence of multiple root-cause categories in a benchmark does not necessarily indicate that its associated detection approach can recognize the underlying mechanisms of each category.


This observation motivates a further investigation into the effectiveness of existing detection approaches.
Although current evaluation datasets cover diverse ESM vulnerability types, whether the underlying detection mechanisms are capable of capturing the corresponding root causes remains unclear. 
Their uneven distributions may obscure detection weaknesses in less-represented categories. Therefore, assessing detection capabilities requires moving beyond aggregate performance metrics to examine effectiveness across individual root causes. We conduct this analysis in RQ3 to identify the detection limitations associated with different vulnerability mechanisms.

\begin{resultbox}
    \textbf{Finding 3:} Existing ESM vulnerability evaluation datasets cover multiple root causes but exhibit uneven distributions, with many causes concentrated in a few manipulation mechanisms. Such imbalances highlight the importance of evaluating detection approaches at the root-cause level rather than relying solely on aggregate detection performance.
\end{resultbox}

\subsection{RQ3: Effectiveness of SOTA Detection Approaches}
\subsubsection{Experiment Setup}
To systematically evaluate the effectiveness of state-of-the-art (SOTA) tools in detecting ESM vulnerabilities, we conduct experiments on both real-world attack incidents and benign DeFi applications. Specifically, for vulnerability detection, we use our dataset of 276 real-world attack incidents to evaluate the detection capability of existing approaches. To evaluate false positives, we further apply these approaches to a benign DeFi application dataset. Specifically, we adopt the benchmark introduced by DeFiScope~\cite{zhong2025detecting}, which contains 441 high-quality DeFi applications collected from DeFiLlama~\cite{DefiLlama}. 

Since the selected approaches rely on different analysis inputs and detection mechanisms, we evaluate them according to their respective requirements.
For approaches that require transaction information, including DeFiRanger, DeFort, and DeFiScope, we follow their original evaluation settings and procedures described in their original studies. Specifically, we reproduce DeFiRanger using a publicly available implementation~\cite{DeFiRanger} and configure it according to the descriptions provided in the original paper. However, since DeFort is close-sourced and DeFiScope does not release its fine-tuned model, we directly report the results from their original papers. For bytecode-based approaches, including DeFiTainter and SmartCat, we perform detection based on the deployed bytecode of relevant contracts. Specifically, we provide the victim contract addresses and attack contract addresses associated with each incident as inputs for evaluation. For LLM-assisted source code analysis, including GPTScan and PMDetector, we collect the source code of the affected victim contracts and apply these approaches for vulnerability detection. We use the same LLM configurations reported in their original papers, i.e., GPT-4o for GPTScan and GPT-4.1 for PMDetector.

\subsubsection{Results}
\begin{figure}[htbp]
  \centering

  \begin{minipage}[b]{0.46\columnwidth}
    \centering
    \scriptsize
    \renewcommand{\arraystretch}{0.92}

    \captionof{table}{Detection results on real attacks and high-quality DeFi applications.}
    \label{tab:tool_detection_outcomes}

    \begin{tabular}{lrrrrrr}
      \toprule
      \multirow{2}{*}{Tool}
      & \multicolumn{3}{c}{Real Attacks}
      & \multicolumn{3}{c}{High-Quality} \\
      \cmidrule(lr){2-4}
      \cmidrule(lr){5-7}
      & Det. & FN. & Fail.
      & FP & Norm. & Fail. \\
      \midrule
      DeFiRanger  & 69 & 202 & 5 & 0 & 441 & 0 \\
      DeFort      & 50 & 45 & -- & 0 & 441 & 0 \\
      DeFiScope   & 76 & 19 & -- & 0 & 441 & 0 \\
      DeFiTainter & 96 & 180 & 0 & 7 & 434 & 0 \\
      SmartCat    & 139 & 101 & 36 & -- & -- & -- \\
      GPTScan     & 134 & 106 & 36 & 54 & 349 & 38 \\
      PMDetector  & 87 & 153 & 36 & 13 & 390 & 38 \\
      \bottomrule
    \end{tabular}

  \end{minipage}
  \hfill
\begin{minipage}[t]{0.50\columnwidth}
    \centering

    \vspace{-85pt}   

    \includegraphics[width=0.99\linewidth]{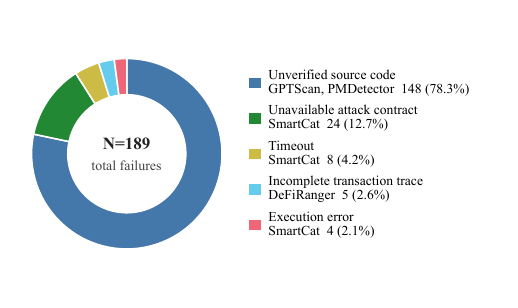}

    \vspace{-20pt}
  \caption{Distribution of failure causes}
  \label{fig:failure-causes}

\end{minipage}

\end{figure}

Table~\ref{tab:tool_detection_outcomes} summarizes the detection effectiveness of SOTA approaches on real-world attacks and benign DeFi applications. For attack cases, we report the numbers of detected vulnerabilities (Det.), false negatives (FN), and analysis failures (Fail.). For benign applications, we report false positives (FP), correctly classified benign cases (Norm.), and analysis failures (Fail.).
Overall, our evaluation reveals substantial limitations in existing approaches for detecting ESM vulnerabilities. Although several approaches can identify specific categories of attacks, none of them achieves comprehensive coverage across the diverse ESM vulnerabilities observed in practice.

Among transaction-dependent approaches, DeFiRanger successfully detects 69 of 276 real-world attacks, while missing 202 cases. Although it produces no false positives on the benign applications, its substantial number of missed attacks indicates limited coverage of ESM behaviors beyond the defined attack patterns,
DeFort and DeFiScope achieve higher detection rates on their evaluated datasets, with recalls of 52.6\% and 80.0\%, respectively, while producing no false positives. More fundamentally, these approaches rely on transaction-level information, which limits their applicability for proactive vulnerability detection before attacks occur. Although pending transactions in the public mempool may provide early signals, attackers can bypass such monitoring through private transaction channels (e.g., Flashbots), making transaction-dependent analysis insufficient for comprehensive ESM prevention.

For bytecode-based approaches, DeFiTainter detects 96 attack incidents with 7 false positives and 180 false negatives. SmartCat achieves the highest detection coverage among the evaluated approaches, identifying 139 attack cases while still missing 101 cases. However, SmartCat primarily focuses on analyzing attack contract and therefore cannot be directly evaluated on benign DeFi applications. LLM-assisted approaches provide enhanced semantic understanding of contract logic but introduce additional challenges. GPTScan detects 135 attack cases, achieving comparable coverage to SmartCat, but produces the highest number of false positives (54 cases) on benign applications. In contrast, PMDetector exhibits more conservative detection behavior, identifying 87 attack cases with 13 false positives.

\begin{resultbox}
    \textbf{Finding 4:} Existing SOTA approaches demonstrate limited effectiveness in detecting real-world ESM vulnerabilities. The best-performing approach detects only 50.4\% of evaluated attack incidents, indicating substantial missed cases in practice. Meanwhile, GPTScan reports the most false positives (54), while other methods produce fewer false positives. These results highlight the challenge of maximizing detection coverage while minimizing false positives.
\end{resultbox}

\noindent \textbf{Failure Case Analysis:} We further examine the analysis failure cases reported in Table~\ref{tab:tool_detection_outcomes}. As shown in Fig.~\ref{fig:failure-causes}, we identify 189 failure cases across different tool-case evaluations, which primarily result from missing analysis artifacts and execution limitations.
The dominant failure cause is the unavailability of verified source code, accounting for 148 cases (78.3\%) from GPTScan and PMDetector. Since both approaches rely on source-level semantic analysis, they cannot process contracts without publicly verified source code. For SmartCat, 36 failures are observed, including 24 cases caused by unavailable attack contracts (e.g., self-destructed attack contracts), 8 cases caused by analysis timeouts, and 4 cases caused by execution errors during analysis. DeFiRanger encounters 5 failures due to incomplete transaction traces, preventing it from reconstructing the required attack behaviors.

\begin{table*}[htbp]
\centering
\caption{Effectiveness of SOTA tools across Root-Cause Taxonomy Categories}
\label{tab:rq4_sota_effectiveness}
\scriptsize
\setlength{\tabcolsep}{2.2pt}
\renewcommand{\arraystretch}{1.15}
\resizebox{\textwidth}{!}{%
\begin{tabular}{lccccccccccc}
\toprule
\textbf{Tool} &
\textbf{RC-A1} & \textbf{RC-A2} & \textbf{RC-A3} &
\textbf{RC-B1} & \textbf{RC-B2} & \textbf{RC-B3} &
\textbf{RC-C1} & \textbf{RC-C2} & \textbf{RC-C3} &
\textbf{RC-D1} & \textbf{RC-D2} \\
\midrule
DeFiRanger  & \rcbar{16}{72} & \rcbar{1}{15} & \rcbar{1}{3} & \rcbar{4}{18} & \rcbar{4}{9} & \rcbar{2}{25} & \rcbar{28}{72} & \rcbar{2}{11} & \rcbar{4}{18} & \rcbar{5}{14} & \rcbar{2}{14}\\
DeFort    & \rcbar{18}{28} & \rcbar{1}{4} &  -- & \rcbar{0}{5} & \rcbar{1}{4} & \rcbar{6}{7} & \rcbar{18}{29} & \rcbar{1}{3} & \rcbar{3}{7} & \rcbar{2}{5} & \rcbar{0}{3} \\
DeFiScope    & \rcbar{20}{28} & \rcbar{2}{4} & -- & \rcbar{4}{5} & \rcbar{3}{4} & \rcbar{6}{7} & \rcbar{25}{29} & \rcbar{3}{3} & \rcbar{6}{7} & \rcbar{4}{5} & \rcbar{3}{3}\\
DeFiTainter & \rcbar{29}{75} & \rcbar{5}{15} & \rcbar{0}{3} & \rcbar{6}{19} & \rcbar{2}{9} & \rcbar{16}{26} & \rcbar{17}{72} & \rcbar{3}{11} & \rcbar{7}{18} & \rcbar{7}{14} & \rcbar{4}{14} \\
SmartCat    & \rcbar{36}{69} & \rcbar{6}{10} & \rcbar{2}{3} & \rcbar{5}{13} & \rcbar{4}{9} & \rcbar{14}{25} & \rcbar{46}{62} & \rcbar{6}{9} & \rcbar{10}{14} & \rcbar{4}{14} & \rcbar{6}{12}\\
GPTScan     & \rcbar{36}{65} & \rcbar{5}{11} & \rcbar{1}{3} & \rcbar{7}{17} & \rcbar{1}{7} & \rcbar{13}{23} & \rcbar{34}{63} & \rcbar{8}{11} & \rcbar{7}{15} & \rcbar{10}{13} & \rcbar{12}{13} \\
PMDetector  & \rcbar{26}{64} & \rcbar{4}{11} & \rcbar{2}{3} & \rcbar{5}{17} & \rcbar{3}{7} & \rcbar{10}{23} & \rcbar{18}{63} & \rcbar{2}{11} & \rcbar{4}{15} & \rcbar{6}{13} & \rcbar{7}{13}  \\
\bottomrule
\end{tabular}%
}
\end{table*}

\noindent\textbf{False Negative Analysis:} To understand the underlying causes of missed detections, we further analyze false negatives from the perspective of our ESM taxonomy. Table~\ref{tab:rq4_sota_effectiveness} reports the detection coverage of each approach across the eleven ESM subcategories, where each entry reports the number of detected cases over the total number of evaluated cases within the corresponding category. Existing approaches exhibit different FN patterns due to their detection strategies. 

(1) \textbf{Transaction-dependent approaches}. DeFiRanger, Defort, and DeFiScope identify price manipulation by analyzing historical transaction traces and observable attack behaviors. Although these approaches are effective when attacks exhibit recognizable execution patterns, their false negatives reveal limitations in capturing ESM vulnerabilities that exploit protocol-specific economic logic without producing the expected behavioral signals.

DeFiRanger reconstructs cash-flow trees from token transfers and matches the recovered behaviors against predefined attack patterns. However, our analysis shows that many missed cases do not exhibit trading patterns. Instead, attackers exploit protocol-specific economic logic, such as incorrect share calculations or exchange-rate manipulation. These attacks may involve legitimate token transfers without forming the cash-flows expected by DeFiRanger, leading to missed detections. 

Defort's detection capability is constrained by its reliance on observable signals, i.e., abnormal token-price changes and attacker profits. This design is effective for conventional price-manipulation attacks. However, it provides limited coverage for broader ESM vulnerabilities, where the manipulated state may not correspond to an explicit price or immediate token profit. For example, the economic impact of an attack may not be realized as direct token profits. Attackers may instead obtain benefits through inflated collateral positions or excessive claims, making such attacks difficult to capture through profit-oriented behavioral analysis. 

DeFiScope faces two core limitations in detecting ESM vulnerabilities. First, it cannot recover the complete DeFi operations from transaction traces. For example, intermediary contracts and non-standard token behaviors may prevent DeFiScope from reconstructing the actual economic action. Second, it matches recovered behaviors against a predefined set of attack strategies, such as buy-and-sell and stake-and-claim patterns. These patterns provide limited coverage for vulnerabilities involving liquidation, internal accounting corruption, or long multi-step state preparation.

\begin{resultbox}
    \textbf{Finding 5:} 
    Transaction-based approaches rely on transaction traces, known attack execution patterns and observable price behaviors. Thus, they may miss attacks that manipulate hidden economic states, and cannot provide effective vulnerability detection before attack execution.
\end{resultbox}

(2) \textbf{Bytecode-based approaches.} DeFiTainter and SmartCat analyze contract bytecode to detect vulnerabilities proactively. DeFiTainter focuses on whether manipulated values propagate to security-critical operations, whereas SmartCat analyzes attack contracts for recognizable manipulation behaviors.

\begin{figure}[htbp]
\centerline{\includegraphics[width=0.92\linewidth]{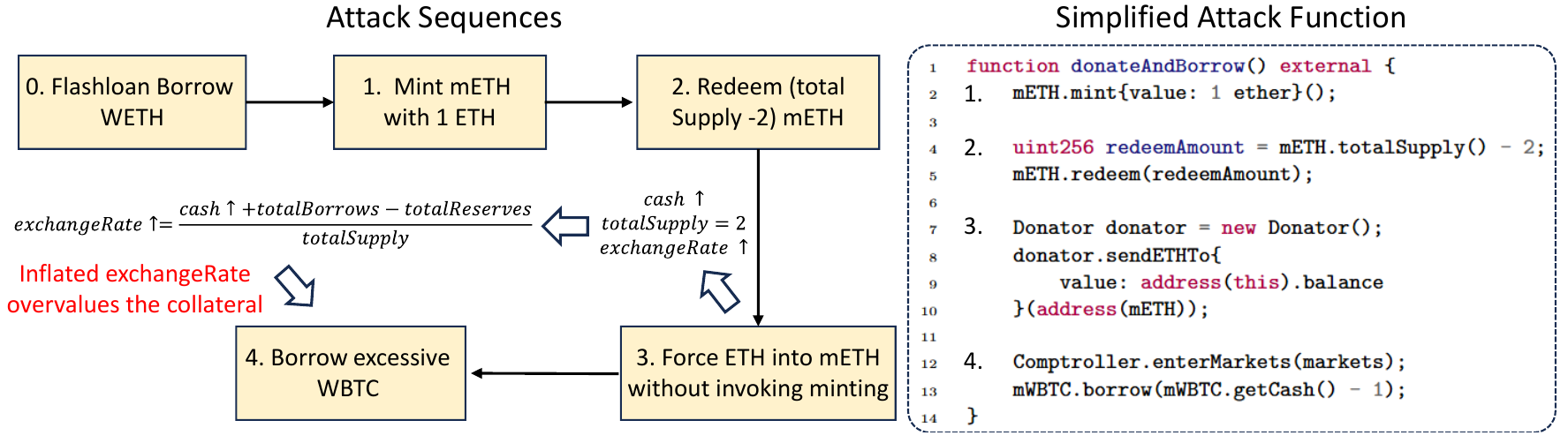}}
\caption{Workflow of the MetaLend Attack}
\label{MetaLend}
\end{figure}

DeFiTainter relies on predefined function signatures to identify potential taint sources and sinks. Table~\ref{tab:rq4_sota_effectiveness} shows that DeFiTainter detects 29 cases in RC-A1 and 16 cases in RC-B3, where vulnerabilities often involve explicit manipulation of market states. However, its signature-based abstraction limits its ability to detect complex manipulation vulnerabilities. For example, shown in Fig.~\ref{MetaLend}, in the MetaLend incident~\cite{MetaLend}, the security-critical sink was not a conventional token transfer operation but an inflated exchange rate maintained through internal accounting rates. Consequently, DeFiTainter failed to identify it. This limitation is also reflected in categories that require reasoning about complex protocol states. For example, DeFiTainter detects only 2 out of 9 RC-B2 cases (22.2\%) and 4 out of 14 RC-D2 cases (28.6\%).

The primary limitation of SmartCat lies in its reliance on predefined manipulation patterns. Its detection rules are mainly designed around DEX/AMM-oriented behaviors. Consequently, attacks targeting other DeFi protocols, such as lending or yield-farming platforms, may not exhibit recognizable patterns and therefore remain challenging to detect. Table~\ref{tab:rq4_sota_effectiveness} shows that SmartCat achieves relatively higher coverage on categories involving explicit token and liquidity manipulation. For example, it detects 74.2\% RC-C1 cases, 66.7\% RC-C2 cases, and 71.4\% RC-C3 cases, where vulnerabilities typically manifest as observable token and liquidity behaviors. In contrast, its coverage decreases for vulnerabilities involving complex protocol-level economic logic, including RC-B1 (38.5\%) and RC-B2 (44.4\%). For example, as shown in Fig.~\ref{MetaLend}, in the MetaLend incident~\cite{MetaLend}, the attacker manipulated the mETH exchange rate by force-sending ETH to the mETH contract and subsequently used the inflated collateral value to borrow excessive WBTC. However, this exploitation sequence does not match SmartCat's predefined DEX/AMM-oriented manipulation patterns.

\begin{resultbox}
    \textbf{Finding 6:} 
    Bytecode-based approaches depend on manually defined taint sources/sinks and behavior patterns, limiting their ability to detect ESM vulnerabilities involving complex contract semantics and protocol-specific economic logic. 
\end{resultbox}

(3) \textbf{LLM-assisted sourcecode approaches.} GPTScan and PMDetector combine sourcecode analysis with LLM-based reasoning to improve semantic understanding of contract logic. GPTScan first identifies candidate function based on predefined vulnerability patterns and then employs LLMs to assess their potential security impact. PMDetector constructs taint paths from potentially manipulated economic inputs to security-sensitive operations and leverages LLMs to access their exploitability.

GPTScan is primarily limited by its rule-guided candidate function selection. Before invoking the LLM, GPTScan filters functions using predefined function names and code-level keywords. Consequently, vulnerabilities implemented with previously unseen identifiers or external interfaces may be excluded before semantic analysis. For example, in the WSM incident~\cite{WSM}, the vulnerable \texttt{fetchPrice} function directly used the DEX \texttt{quoteExactOutputSingle} result to determine the presale token price, making the calculation dependent on manipulable market states. Although GPTScan successfully analyzed the contract and extracted 13 functions, neither \texttt{fetchPrice} nor \texttt{quoteExactOutputSingle} matched the predefined price-manipulation rule. As a result, no candidate function was submitted to the LLM, leading to the missed detection. Table~\ref{tab:rq4_sota_effectiveness} reflects this limitation. GPTScan detects only 41.2\% RC-B1 cases and 14.3\% RC-B2 cases, where vulnerabilities often involve complex accounting relationships.

PMDetector is primarily constrained by its sourece-to-sink taint abstraction. It can identify vulnerabilities only when manipulated values propagate from predefined taint sources to recognized security-critical sinks. If no relevant taint path is constructed, the contract is excluded from subsequent LLM-based reasoning. This limitation is reflected in its low coverage of token and AMM related vulnerabilities: PMDetector detects only 17/63 RC-C1 cases (27.0\%), 2/11 RC-C2 cases (18.2\%), and 4/15 RC-C3 cases (26.7\%). For example, in the LUME incident~\cite{LUME}, the public \texttt{burn} function allowed any callers to burn token without verifying the \texttt{msg.sender}. The attacker therefore reduced the AMM pair token balance and distorted the pool state. However, PMDetector did not recognize the compound storage updates to \texttt{balances[from]} as security-critical sinks. Consequently, it constructed no taint path, preventing the vulnerable function from reaching the LLM analysis stage and resulting in a missed detection.


\begin{resultbox}
    \textbf{Finding 7:} 
 Existing approaches mainly use LLMs as a post-analysis reasoning component, where LLMs validate candidates generated by rule-based filtering or taint-based analysis. Consequently, their detection coverage remains constrained by the completeness of the preceding static analysis abstractions, causing vulnerabilities outside predefined patterns or analysis paths to be missed.
\end{resultbox}

\noindent\textbf{False Positive Analysis:}
To understand why existing approaches report vulnerabilities in benign DeFi applications, two authors independently manually inspected all false positives generated by these methods to verify their validity and identify the underlying causes. The results are summarized in Table~\ref{tab:tool_detection_outcomes}. In our evaluation, the transaction-dependent approaches produce no false positives on the high-quality applications. These approaches rely on concrete execution evidence, such as abnormal asset flows and observable price changes. Since benign DeFi applications typically perform legitimate operations that do not exhibit such coordinated manipulation behaviors, these evidence-driven approaches don't generate false alarms in our benchmark. However, this conservative detection strategy also limits their ability to detect vulnerabilities proactively.

DeFiTainter's false positives mainly arise from interpreting value propagation as evidence of exploitable manipulation. Four of its seven false positives are caused by missing authorization and execution semantics. For example, the reported tainted path in FFIST~\cite{Benign} originates from standard \texttt{transferFrom} function protected by allowance checks. The remaining cases result from the taint abstraction, which treats externally influenced asset transfers as suspicious without reasoning about authorization constraints or economic invariants. For example, the \texttt{swap} function in LeetSwap protocol~\cite{Benign}, which follows a standard AMM design with reserve constraints and invariant checks, is flagged.

GPTScan produces the largest number of false positives, with 54 applications generating vulnerability reports. Our analysis shows that these false positives mainly arise from over-generalizing rule-guided vulnerability patterns without reconstructing their complete protocol context. Specifically, 16 applications are misclassified due to missing authorization or caller-context semantics. For example, it flagged \texttt{wrapper} function in Acoconut~\cite{Benign}, which delegates the transfer to an inherited \texttt{transferFrom} implementation. However, the contract enforces the allowance checks. Another 11 cases misinterpret internal liquidity operations as user-directed swaps without slippage protection. The remaining cases arise from analyzing auxiliary contracts as core protocol logic, misinterpreting checkpoint mechanisms, or making unverified assumptions about external dependencies.


PMDetector's false positives mainly arise from confusing potential data dependencies with economically exploitable manipulation. Seven of its false positives are caused by unsupported assumptions about external dependencies. Although PMDetector identifies data flows involving oracle values or market indices, the subsequent analysis does not always establish whether these values can realistically be controlled by attackers. Other false positives result from interpreting legitimate protocol mechanisms as attacks or overstimating the exploitability of feasible taint paths. For example, some reported taint paths demonstrate possible data influence but fail to establish victim asset impact or attacker profit. These cases indicate that PMDetector’s taint analysis can identify plausible dependency chains, but additional economic reasoning is required to distinguish potential influence from practical exploitation.

\begin{resultbox}
    \textbf{Finding 8:} False positives in proactive approaches mainly arise from incomplete exploitability validation. Existing methods can identify suspicious code patterns or potential data dependencies, but often fail to verify whether the corresponding state is actually attacker-controllable or whether the execution is permitted under authorization and protocol constraints.
\end{resultbox}

\section{Discussion}

\subsection{Implications}

\noindent \textbf{For DeFi developers.} Based on our findings, we suggest the following improvements for DeFi protocol developers: \ding{172} \textit{Treat economic-state integrity as a core security requirement.} Developers should not only focus on explicit price inputs or oracle security. Our taxonomy shows that ESM vulnerabilities can also arise from protocol-internal accounting and AMM and liquidity states. Therefore, the integrity of economically sensitive states and their underlying assumptions should be explicitly modeled and validated during protocol design. \ding{173} \textit{Examine economic dependencies across protocol components.} ESM vulnerabilities often emerge from interactions among multiple protocol components, including token behaviors, AMM and liquidity mechanisms, and higher-level business logic. Developers should avoid evaluating contracts or functions in isolation and instead examine how state changes in one component propagate to others.  In particular, token-level and liquidity-state changes should be traced to determine whether they can lead to incorrect valuation, accounting, or other security-critical decisions. \ding{174} \textit{Combine complementary security analysis before deployment.} Our evaluation shows that no existing detector provides comprehensive coverage of ESM vulnerabilities, with the best-performing approach detecting only 50.4\% of evaluated real-world attacks. Developers should therefore not treat any single automated tool as sufficient evidence of security. Instead, pre-deployment security assessment should combine complementary automated analysis with manual review and dynamic analysis.

\noindent \textbf{For security practitioners.} For security practitioners working at developing ESM vulnerability detection tools, we provide the following implications for improving the analysis of ESM vulnerabilities. 
\ding{172} \textit{Focus on violated economic assumptions rathe than observable symptoms.} Future tools should look beyond abnormal prices, or suspicious cash flows, and instead identify which economic state is affected and which integrity assumption is violated. As our taxonomy shows, similar economic consequences may arise from fundamentally different root causes. Reasoning about economic-state integrity can therefore help uncover vulnerabilities that do not exhibit conventional patterns.
\ding{173} \textit{Broaden detection rules to cover overlooked manipulation mechanisms.} Existing approaches are often designed around a limited set of predefined attack patterns. Our taxonomy provides a systematic characterization of manipulated economic states and violated assumptions, which can be further translated into detection rules. Incorporating these taxonomy-derived rules can help extend the coverage of existing tools to previously overlooked manipulation scenarios.
\ding{174} \textit{Strengthen reachability and exploitability analysis.} Detecting suspicious code patterns or data dependencies alone is insufficient to establish an exploitable vulnerability. Future approaches should further determine whether the affected state is attacker-controllable, whether the vulnerable execution path is reachable, and whether the resulting state violation can produce a meaningful economic impact. Combining program analysis with LLM-based semantic reasoning may provide a promising direction for improving both exploitability assessment and the interpretability of detection results.

\subsection{Future Directions} Our empirical findings reveal several fundamental limitations in existing ESM detection approaches. These observations suggest several promising directions for future research. First, future approaches should
move beyond price-centric detection toward systematic reasoning about economic-state integrity. Our evaluation shows that existing approaches remain constrained by predefined patterns and taint abstractions, even when LLMs are incorporated for semantic analysis. A promising direction is to more effectively integrate LLM-based semantic reasoning with program analysis, incorporating semantic reasoning earlier in the vulnerability discovery process rather than using LLMs solely to validate preselected candidates. Specifically, LLMs could help identify economically sensitive operations, infer protocol-level economic constraints, and discover previously unrecognized manipulation mechanisms, while program analysis handles deterministic tasks, such as dependency tracing and control-flow analysis, to verify the identified relationships. Furthermore, our false-positive analysis highlights the challenge of determining whether suspicious economic-state dependencies are actually exploitable. Future approaches could incorporate stateful simulation to examine whether attackers can feasibly manipulate economic states, violate integrity constraints, and exploit these violations through security-critical operations. 

\subsection{Threats to Validity}

\noindent\textbf{External Validity:}
A primary threat to validity concerns the representativeness of our dataset. Since publicly documented attack incidents may be concentrated on common or well-understood attack patterns, our findings could be biased toward certain vulnerability types. To mitigate this threat, we systematically collected incident records from DeFiHackLabs, incorporated publicly available incidents from existing PM/ESM datasets, and examined Code4rena audit reports to capture both exploited attacks and proactively discovered vulnerabilities. These complementary sources help improve the diversity and coverage of our dataset. Nevertheless, undocumented attacks and vulnerabilities in emerging DeFi protocols may not be fully represented, potentially limiting the generalizability of our findings. Another threat arises from the selection of evaluation subjects. Our false-positive evaluation relies on high-quality DeFi applications, which may not fully represent the diversity of deployed DeFi contracts.

\noindent\textbf{Internal Validity:} A potential threat to internal validity arises from subjective judgments during vulnerability classification and taxonomy construction. Different researchers may interpret the underlying root causes of complex ESM vulnerabilities differently, potentially introducing classification bias. To mitigate this threat, two researchers independently analyzed and categorized the collected cases using open card sorting, with a third experienced research adjudicating unresolved disagreements. Another threat concerns the reproducibility and consistency of tool evaluations, as different approaches require different analysis inputs and execution environments. To minimize this threat, we followed the original evaluation settings and configurations whenever possible. For transient failures caused by LLM API or RPC request errors, we identified and reran the affected cases under the same configuration.

\section{Related Work}

\subsection{Price Manipulation Vulnerability Detection}

PM vulnerabilities are critical security threats in DeFi, attracting extensive research efforts~\cite{xi2024pomabuster,zhang2026spsyn,pratiwi2026defitrace,deng2024safeguarding,heimbach2024non,mongardini2026midsummer,ahmadjee2025decision}. DeFiRanger~\cite{wu2023defiranger} constructs cash flow trees and lifts low-level transaction semantics into high-level DeFi actions, enabling the identification of PM attack. Similarly, DeFiGuard~\cite{DeFiGuard} transforms raw transaction data into cash-flow graphs and applies graph neural networks to detect anomalous patterns associated with PM attacks. Beyond transaction-based methods, DeFiTainter~\cite{kong2023defitainter} employs rule-based taint analysis to identify PM vulnerabilities, while SmartCat~\cite{zhang2025following} further advances this line of work by analyzing deployed contracts in real time, detecting PM attacks from the perspective of attack contracts. More recently, the emergence of large language models has inspired new directions in vulnerability detection. GPTScan~\cite{sun2024gptscan} integrates LLMs with static analysis, enabling more expressive detection of PM vulnerabilities. DeFiScope~\cite{zhong2025detecting} proposes an LLM-driven approach that infers potential PM attacks by predicting price fluctuation trends.

\subsection{Empirical Study on Vulnerability Detection for Smart Contracts}

As the primary carriers of digital assets in DeFi, smart contracts have attracted significant attention. Many studies conducted empirical evaluations of existing vulnerability detection tools to assess their effectiveness~\cite{durieux2020empirical,perez2021smart,bodell2023proxy,zhang2025empirical,bobadilla2025automated,chen2025characterizing,wang2026empirical,pacheco2023my}. Ferreira et al.~\cite{ferreira2020smartbugs} presented SmartBugs, an extensible execution framework that evaluates ten analysis tools over 47,661 real-world contracts. Chaliasos et al.~\cite{chaliasos2024smart} further assessed five SOTA security tools on 127 real-world attacks, highlighting the necessity for specialized tools tailored to diverse vulnerability scenarios. Moreover, several studies focus on systematic evaluations of static analysis and fuzzing techniques~\cite{li2024static,wu2024we,sharma2022survey,chen2025chatgpt}. Ghaleb et al.~\cite{ghaleb2020effective} proposed SolidFi, an automated and systematic framework for evaluating static analysis tools for smart contracts. And Wu et al.~\cite{wu2024we} constructed a benchmark of 2,000 carefully labeled contracts and conducted a large-scale evaluation of 11 state-of-the-art smart contract fuzzers. Beyond general-purpose evaluations, some studies investigate specific vulnerability classes or contract ecosystems~\cite{sun2024all,huang2026revealing,zhang2024contracts,wang2026one,ma2025comprehensive}. Song et al.~\cite{song2024empirical} conducted an empirical study on Move-based smart contracts, identifying eight categories of defects. Liu et al.~\cite{liu2025have} focused on access control vulnerabilities, evaluating six detection tools and uncovering their limitations in identifying real-world access control flaws.

\section{Conclusion}

This paper presents a comprehensive empirical study of Economic-State Manipulation (ESM) vulnerabilities in DeFi. We construct a multi-source dataset comprising 276 real-world attacks and 91 audit findings, based on which we develop a cause-oriented taxonomy covering four primary categories and eleven subcategories. Our analysis reveals that ESM vulnerabilities extend beyond conventional price manipulation, often arising from violations of underlying economic-state integrity assumptions. Through a systematic evaluation of seven representative detection approaches, we find that existing techniques provide limited coverage, with the best-performing approach detecting only 50.4\% of real-world attacks. Our analysis attributes these limitations to reliance on predefined attack patterns, incomplete understanding of economic semantics, and insufficient exploitability validation. These findings highlight the need for more generalizable, economic-state-aware detection techniques and provide empirical insights and guidance to DeFi protocol developers and security practitioners.

\bibliographystyle{ACM-Reference-Format}
\bibliography{ESM}










\end{document}